\documentclass[twocolumn, aps, superscriptaddress, shortbibliography, showpacs,amsmath,amssymb,floatfix, english]{revtex4-1}
\usepackage{lipsum}
\usepackage{graphicx}% Include figure files
\usepackage{dcolumn}% Align table columns on decimal point
\usepackage{amsthm}
\usepackage{bm}% bold math

\pdfoutput=1
\usepackage{graphicx,graphics,epsfig,subfigure,times,bm,bbm,amssymb,amsmath,amsthm,mathrsfs,MnSymbol}
\usepackage{gensymb}
\usepackage{amsfonts}
\usepackage[matrix,frame,arrow]{xypic}
\usepackage[pdfstartview=FitH]{hyperref}
\hypersetup{
    colorlinks=true,       % false: boxed links; true: colored links
    linkcolor=red,          % color of internal links
    citecolor=magenta,        % color of links to bibliography
    filecolor=magenta,      % color of file links
    urlcolor=cyan,           % color of external links
    runcolor=cyan}
\usepackage{epstopdf}
\usepackage[pdftex]{color}
\usepackage{hypernat}
\usepackage{braket}%Dirac Notation in QM
\usepackage{enumerate}
\usepackage[normalem]{ulem}
\usepackage[usenames,dvipsnames]{xcolor}
\usepackage{multirow}
\usepackage{mathtools}

\definecolor{orange}{rgb}{1,0.5,0}

\newcommand{\ignore}[1]{}

\ignore{
\documentclass[eprintnumbers,amsmath,amssymb,onecolumn,a4paper,caption ]{article}
\usepackage{amsfonts}
\usepackage{amssymb}
\usepackage{mathrsfs}
\usepackage{mathbbold}
\usepackage{bbm}
\usepackage{mathrsfs}
\usepackage{dcolumn}% Align table columns on decimal point
\usepackage{bm}% bold math
\usepackage{times,epsfig,amssymb,amsmath}
\usepackage{float}
\usepackage{subfigure}
\usepackage{geometry}\geometry{left=2.5cm,right=2.5cm,top=3cm,bottom=3cm}

\usepackage{color}

}

\begin{document}

\title{Out-of-time-order correlators and quantum phase transitions in the Rabi and Dicke models}
%Out-of-time-order correlation functions as indicators of quantum phase transitions in the Rabi and Dicke model
%Quantum phase transitions in the Rabi and Dicke model revealed via out-of-time-order correlation functions
%Out-of-time-order correlation functions and quantum phase transitions in the Rabi and Dicke model
\author{Zheng-Hang Sun}
\thanks{Those authors contributed equally to this work.}
\affiliation{Institute of Physics, Chinese Academy of Sciences, Beijing 100190, China}
\affiliation{School of Physical Sciences, University of Chinese Academy of Sciences, Beijing 100190, China}

\author{Jia-Qi Cai}
\thanks{Those authors contributed equally to this work.}
\affiliation{School of Physics, Huazhong University of Science and Technology, Wuhan 430074, China}

\author{Qi-Cheng Tang}
\affiliation{School of Science, Westlake University, Hangzhou 310024, China}

\author{Yong Hu}
\email{huyong@hust.edu.cn}
\affiliation{School of Physics, Huazhong University of Science and Technology, Wuhan 430074, China}

\author{Heng Fan}
\email{hfan@iphy.ac.cn}
\affiliation{Institute of Physics, Chinese Academy of Sciences, Beijing 100190, China}
\affiliation{School of Physical Sciences, University of Chinese Academy of Sciences, Beijing 100190, China}
\affiliation{CAS Central of Excellence in Topological Quantum Computation, Beijing 100190, China}

\begin{abstract}
We use the out-of-time-order correlators (OTOCs) to study the quantum phase transitions (QPTs) between the normal phase and superradiant phase in the Rabi and few-body Dicke models with large frequency ratio of the atomic level splitting to the single-mode electromagnetic radiation field frequency. We focus on the OTOC thermally averaged with infinite temperature, which is an experimentally feasible quantity. We show that the critical points can be identified by long-time averaging of the OTOC via observing its local minimum behavior. More importantly, the scaling laws of the OTOC for QPTs are revealed by studying the experimentally accessible conditions with finite frequency ratio and finite number of atoms in the studied models. The critical exponents extracted from the scaling laws of OTOC indicate that the QPTs in the Rabi and Dicke model belong to the same universality class.
\end{abstract}
%\pacs{Valid PACS appear here}
\maketitle

\section{Introduction}

The equilibrium phases of matter and quantum phase transitions (QPTs) are conventionally understood
by broken symmetries~\cite{book_QPT1} and their topologies~\cite{book_QPT2}.
In regarding of the rapid development in quantum technologies, the nonequilibrium properties of many-body
quantum systems can be studied by large-scale quantum simulators~\cite{QS_a1,QS_a2}.
The dynamics of several quantities, such as the quantum correlations~\cite{dy1} and
order parameters~\cite{QS_a2,dy2,dy3}, play an important role in characterizing the quantum criticality and
may serve as a bridge from the equilibrium phases to the nonequilibrium behaviors of systems.

Recently, the out-of-time-order correlators (OTOCs)~\cite{def1,def2,def3}, which quantify the
quantum chaos~\cite{chaos1,chaos2,chaos3,chaos4},
information scrambling~\cite{scrambling1,scrambling2,scrambling3,scrambling4,scrambling5,scrambling6,scrambling7,scrambling8},
and many-body localization~\cite{mbl1,mbl2,mbl3,mbl4,mbl5,mbl6},
have also been exploited in detecting
the QPTs. The Lyapunov exponent obtained
from the dynamical behavior of OTOC with thermal average
has a maximum near the quantum critical region~\cite{QPT1,QPT2}.
In addition, it has been shown that the OTOCs with respect to
the ground states and quenched states can diagnose the QPTs
and dynamical phase transition respectively~\cite{QPT3}.
Meanwhile, great progresses have been achieved for experimental
observation of the OTOCs by employing inverse-time evolution~\cite{exp1,exp2}.

The Dicke and Rabi models, which are fundamental models in quantum optics,
reveal a universality class of QPT from largely unexcited normal phase to superradiant phase
in the case of
the thermodynamic limit~\cite{Dicke2,Dicke1,Dicke_add1,Dicke_add2,Dicke_add3}
and infinite atomic level splitting in
the unit of the single-mode light field frequency~\cite{Rabi1,Rabi2,Rabi3,Rabi4,Rabi5} respectively.
The critical exponents obtained from the scaling laws of corresponding observables for
the ground states in the Rabi and Dicke models are found to be the same~\cite{Rabi1,Rabi2,critical_exponents1,critical_exponents2}.
This fact implies that there exists a wide class of observables whose scaling behaviors associated with the finite-size and finite-frequency effect can be observed near the critical points of these models.

Although the OTOCs are closely related to the QPTs in condensed-matter systems, it is not clear whether the OTOCs can detect the QPT between the normal phase and the superradiant phase in cavity-atom interaction systems. The Rabi and Dicke models can be experimentally realized in the ultracold atoms~\cite{Dicke_exp1,Dicke_exp_add1,Dicke_exp2,Dicke_exp3}, trapped ions~\cite{Rabi_exp1,Rabi_exp2}, the superconducting circuits~\cite{Rabi_exp3,Rabi_exp4}, as well as the photonic system~\cite{Rabi_exp5}, where the time evolutions of the observables are naturally accessible. The QPTs between normal phase and supperradiant phase in the Dicke and Rabi models are observed via adiabatic evolution of the order parameter~\cite{Dicke_exp1,Dicke_exp_add1,Rabi_exp2}. Additionally, we recognize that the quench dynamics~\cite{Rabi_exp1} or the initialization into thermal equilibrium states~\cite{exp2} are also accessible in some experimental platforms.
On the other hand, the universality and scaling exponents play important roles for our comprehension of QPTs~\cite{book_QPT1}. Consequently, it is important to develop the OTOCs method as a new dynamical way of characterizing the QPTs besides the adiabatic protocol, and clarify whether the scaling laws can be obtained from the behaviors of OTOCs close to the critical points.

In this work, we show that: (\romannumeral1) the OTOCs can detect the QPT between normal phase and superradiant phase in the Rabi model and the Dicke model with small number of atoms. Specifically, taking the experimental accessibility into consideration, we mainly study the OTOC with infinite temperature thermal average. We find that the local minimum point of time-averaged infinite-temperature OTOC coincides with the critical point.
Therefore, it has the potential to be a dynamical probe of QPTs. (\romannumeral2) Our numerical results also reveal several scaling laws of the studied OTOC,
paving the way of extracting significant information of the QPTs in few body systems, such as the universality class and
the location of critical point in the thermodynamic limit or large frequency ratio limit. In short, our results provide an experimentally feasible approach to detect equilibrium quantum critical points and the universality of QPTs.

\section{Results}

\subsection{\label{sec:level2}Definition}

The OTOCs are defined as~\cite{def1,def2,def3}
\begin{eqnarray}
\mathcal{F}(t) = \langle \hat{W}^{\dagger}(t)\hat{V}^{\dagger}(0)\hat{W}(t)\hat{V}(0) \rangle,
\label{def_otoc}
\end{eqnarray}
with two commuting Hermitian operators at equal time, i.e,  $[\hat{W}(0),\hat{V}(0)]=0$, and $\hat{W}(t)=e^{i\hat{H}t}\hat{W}(0)e^{-i\hat{H}t}$.
The average $\langle\mathcal{O}\rangle$ above can be chosen as the thermal average
$\langle\mathcal{O}\rangle_{\beta}=\text{Tr}(e^{-\beta H}\mathcal{O})/ \text{Tr}(e^{-\beta H})$ with the inverse temperature $\beta=1/T$,
or the state average $\langle\mathcal{O}\rangle_{|\psi\rangle}=\langle\psi| \mathcal{O}|\psi\rangle$ with a given pure state $|\psi\rangle$.

\subsection{\label{sec:level2}OTOCs in the Rabi model}
\begin{figure}
  \centering
  \includegraphics[width=0.9\linewidth]{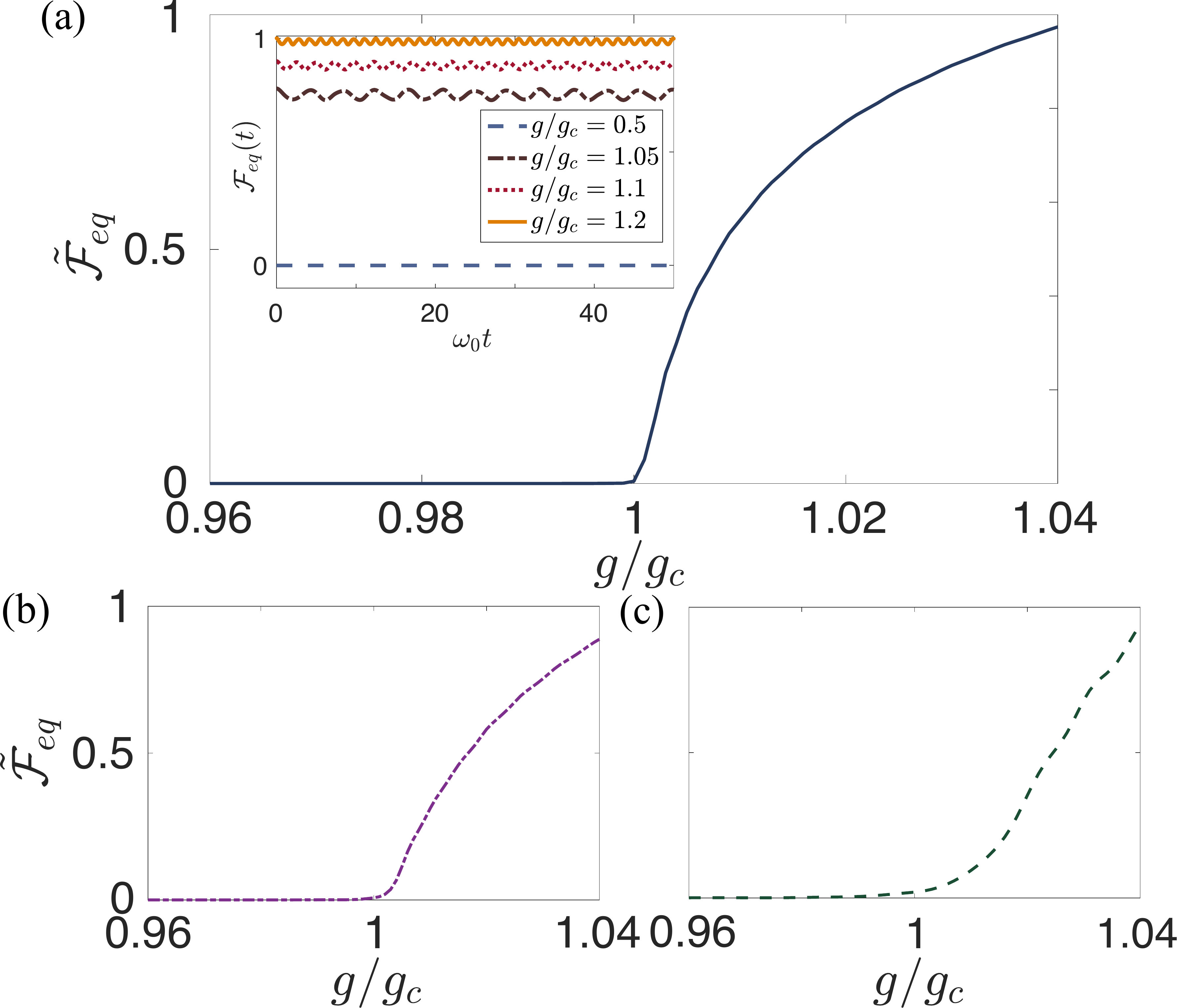}\\
  \caption{(a) The dependence of time-averaged equilibrium OTOC $\tilde{\mathcal{F}}_{eq}$ and $g/g_{c}$. The time window of the average is $t_{f}\sim 500 \omega_{0}$ and the frequency ratio in the Rabi model is $\eta = 2^{20}$. Here, the number of cavity photons is $n=80$. Inset: The time evolution of equilibrium OTOC $\mathcal{F}_{eq}(t)$. (b) and (c) are similar to (a) but for the Rabi model with $n=40$ and $n=20$, respectively. The values of $\tilde{\mathcal{F}}_{eq}$ are rescaled by the $\tilde{\mathcal{F}}_{eq}$ at $g/g_{c}=1.2$.}\label{eq_OTOC}
\end{figure}

We firstly study the OTOC in the Rabi model with Hamiltonian as:
\begin{eqnarray}
\mathcal{H}_{\text{Rabi}} = \omega_{0}a^{\dagger}a+\frac{\Omega}{2} \sigma_{z}+g(a^{\dagger}+a)\sigma_{x},
\label{hamiltonian1}
\end{eqnarray}
where $\omega_{0}$ denotes the frequency of the single-mode light field, and the level splitting of a single two-level atom is $\Omega$,
$g$ denotes the coupling strength between the light field and the atom. The critical coupling strength of the ground state QPT is $g_{c}=\sqrt{\omega_{0}\Omega}/2$, when the QPT happens in the limit of frequency ratio $\eta = \Omega/\omega_{0}\rightarrow \infty$~\cite{Rabi1,Rabi2}. Here we emphasize that when we discuss the QPTs in the Rabi model, although strictly speaking, the limit $\eta\rightarrow\infty$ is not the thermodynamic limit, it plays a similar rule of thermodynamic limit. Hence, we adopt relatively large values of $\eta=2^{11}, 2^{12}, \ldots, 2^{20}$ to investigate the characterization of QPT in the Rabi model via infinite-temperature OTOC and related scaling laws. The quantum Rabi model explicitly breaks the parity symmetry $\Pi=\exp\{i\pi(a^{\dagger}a+\frac{1+\sigma_{z}}{2})\}$. Thus, the QPT of this model can be understood by symmetry broken theory~\cite{book_QPT1}. The system breaks $\Pi$ when $g\geq g_c$ and recovers its symmetry when $g \le g_{c}$~\cite{Rabi1}.

We choose the operators in Eq. (\ref{def_otoc}) $\hat{W}=\hat{V}=a^{\dagger}a$ as the order parameter of this model to define the OTOCs. Here, we realize that the OTOCs may provide non-trivial information of quench dynamics because the eigenstates of $\mathcal{H}_{\text{Rabi}}$ are not necessarily the eigenstates of $a^{\dagger}a$. The exact analytical solution of quenched dynamics instead of adiabatic evolution for the Rabi model is still a complex open problem, and we would numerically study the OTOCs Eq. (\ref{def_otoc}) in the considered models based on exact diagonalization method (the functions "propagator" and "Q.groundstate()" in QuTiP are used to get $U(t)=e^{-i\hat{H}t}$ and the ground state $|\psi\rangle$ of $\hat{H}$, and the thermal average $\langle\mathcal{O}\rangle_{\beta}$ can also be calculated when we obatin all the eigenenergies and eigenstates of $\hat{H}$ using the functions "Q.eigenenergies()" and "Q.eigenstates()" in QuTiP~\cite{qutip}). The python codes used to generate our results are presented in the Supplementary Materials~\cite{SM}. We believe that the following numerical results are valuable for the reference of further analytical works.

\begin{figure}
	\centering
	\includegraphics[width=0.8\linewidth]{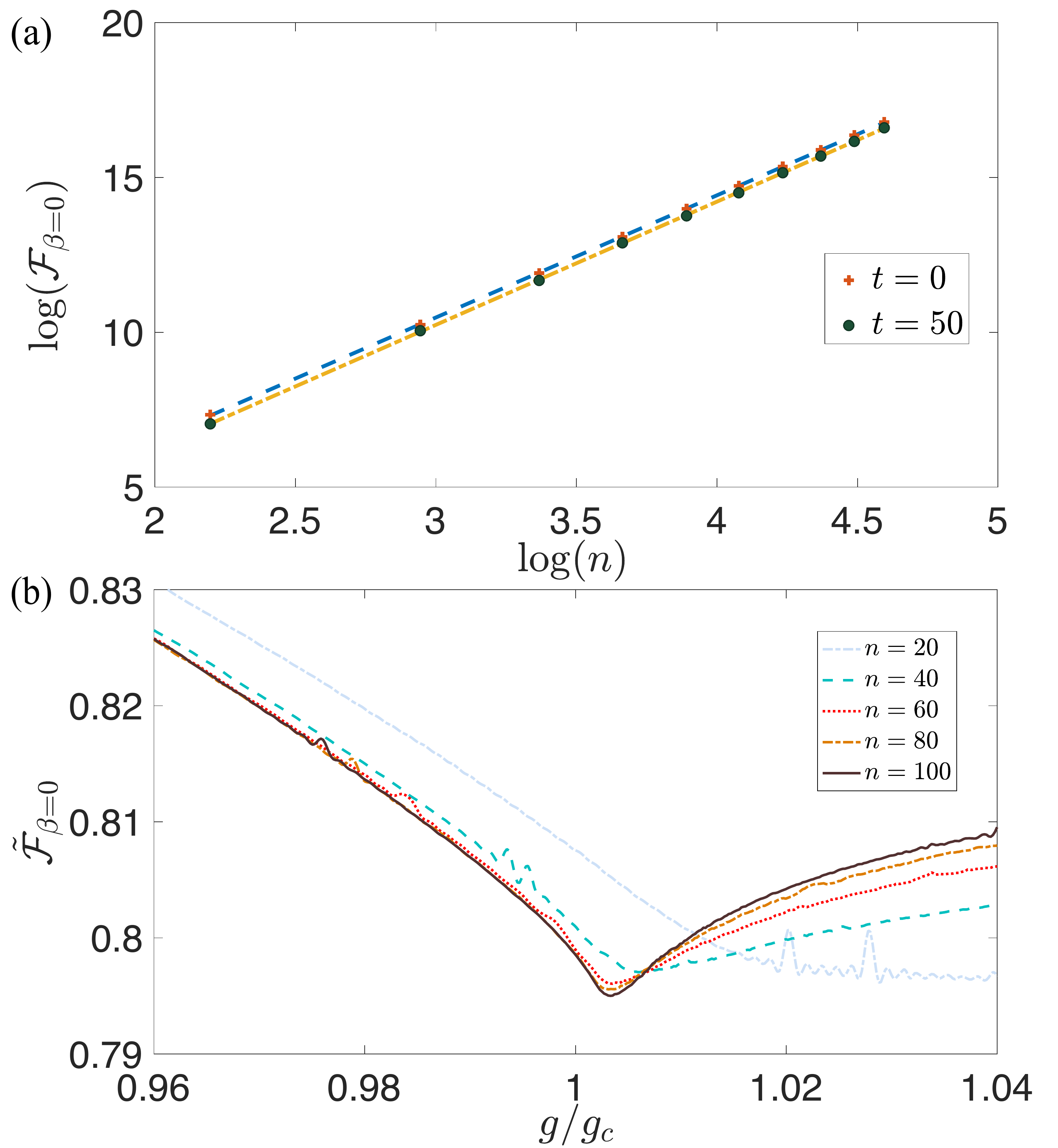}\\
	\caption{(a) The $\log(\mathcal{F}_{\beta=0})$ as a function of $\log(n)$ for time $t=0$ and $t=50$, where $\mathcal{F}_{\beta=0}$ is the bare value of OTOC and $n$ is the maximum photon cutoff. The fitting curves for $t=0$ and $t=50$ are $\log(\mathcal{F}_{\beta=0}) \propto 3.98 \log(n)$ and $\log(\mathcal{F}_{\beta=0}) \propto 3.95 \log(n)$ respectively, which indicates that $\mathcal{F}_{\beta=0}(t) \sim n^{4}$. (b) The time-averaged infinite-temperature OTOC $\tilde{\mathcal{F}}_{\beta=0}$ as a function of $g/g_{c}$ in the Rabi model with $\eta=2^{20}$ and different dimension of operator $a^{\dagger}a$ $n=20,40,60,80,100$.}\label{photon}
\end{figure}

In Fig.~\ref{eq_OTOC} (a), the time-averaged equilibrium OTOC $\tilde{\mathcal{F}}_{eq} = \frac{1}{t_{f}}\int_{0}^{t_{f}}dt \mathcal{F}_{eq}(t)$ as a function of $g/g_{c}$ is presented, which indicates that $\tilde{\mathcal{F}}_{eq}$ can detect the QPT in Rabi model.
In fact, it is obvious that the value of $\mathcal{F}_{eq}(t)$ at any time $t$ can highlight the critical point. We then study the dependence of $\mathcal{F}_{eq}(t)$ and the cutoff on the number of cavity photons $n$ (also as the dimension of the operator $a^{\dagger}a$). The results for $n=40$ and $n=20$ are shown in Fig.~\ref{eq_OTOC} (b) and (c) respectively, while in Fig.~\ref{eq_OTOC} (a), $n=80$ as relatively large value. One can see that the signature of QPT diagnosed by the equilibrium OTOC becomes less distinguishable for small $n$. Thus, 
in the following study, we will choose a large value of $n$ and benchmark the finite-$n$ effect (see Fig. \ref{photon} for the other type of OTOC.)

\begin{figure}
	\centering
	\includegraphics[width=0.9\linewidth]{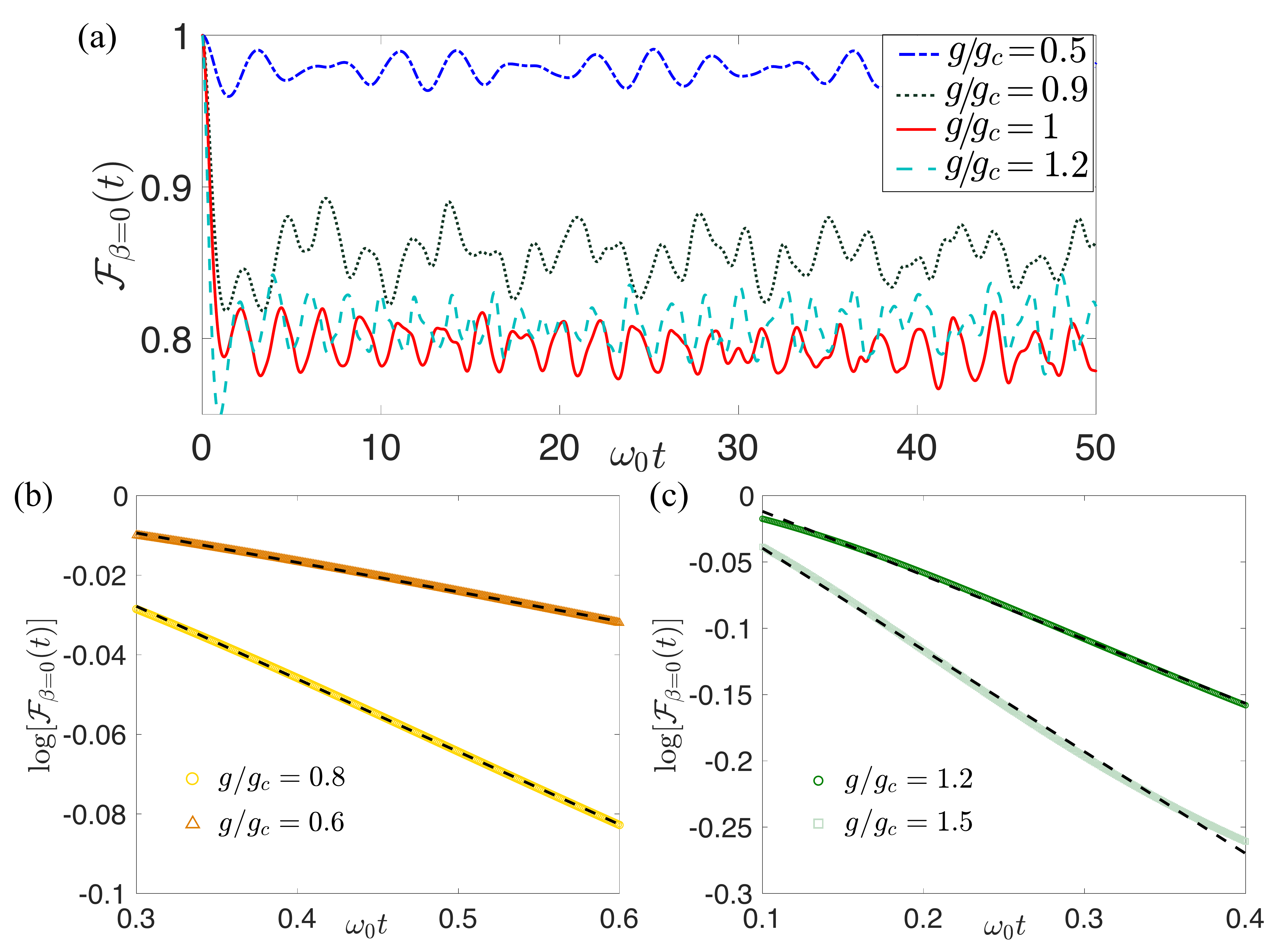}\\
	\caption{(a) The time evolution of infinite-temperature OTOC $\mathcal{F}_{\beta=0}(t)$ (normalized). (b) The dynamical behaviors of $\log(\mathcal{F}_{\beta=0}(t))$ in Rabi model with $g/g_{c} = 0.8,0.6$ and time interval $t\in[0.3\omega_{0},0.6\omega_{0}]$. The dashed lines are linear fittings. (c) Similar to but with $g/g_{c}=1.2,1.5$ and time interval $t\in[1.2\omega_{0},1.5\omega_{0}]$.}\label{inf_OTOC}
\end{figure}
\begin{figure}
	\centering
	\includegraphics[width=0.9\linewidth]{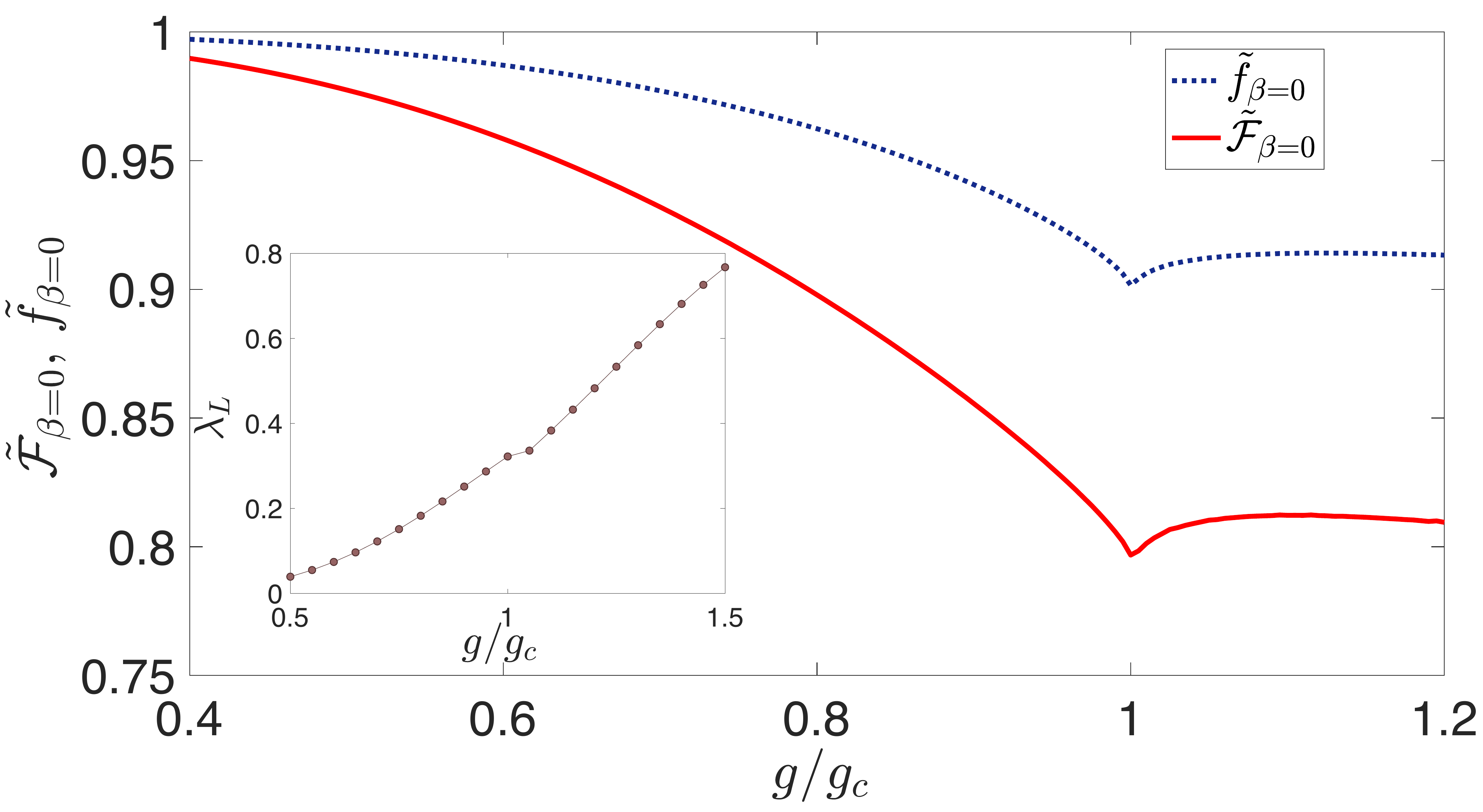}\\
	\caption{The dependence of time-averaged infinite-temperature OTOC $\tilde{\mathcal{F}}_{\beta=0}$ (normalized) as well as two-point correlator $\tilde{f}_{\beta=0}(t)$ (normalized), and $g/g_{c}$. The time window of the average is $t_{f}\sim 500 \omega_{0}$ and the frequency ratio in the Rabi model is $\eta = 2^{20}$. The inset shows the Lyapunov-like exponent extracted from the time evolution of $\mathcal{F}_{\beta=0}(t)$ as a function of $g/g_{c}$.}\label{inf_OTOC_3}
\end{figure}

Nevertheless, efficiently acquiring the ground state of a system, as a necessity for evaluating the equilibrium OTOC~\cite{exp1},
is still a challenge in several experimental platforms~\cite{hard_ground,QS}.
Fortunately, the OTOC thermally averaged with $\beta=0$
$\mathcal{F}_{\beta=0}(t) = \text{Tr}[ \hat{W}^{\dagger}(t)\hat{V}^{\dagger}(0)\hat{W}(t)\hat{V}(0)] $ (the partition function $Z=\text{Tr}(e^{-\beta H})$ with $\beta=0$ is independent of the coupling strength $g$ and thus is omitted)
can be measured in the NMR quantum simulator~\cite{exp2} by applying the distinguishability protocol~\cite{scrambling2}. Here, the quantity
$\mathcal{F}_{\beta=0}(t)$ is named after infinite-temperature OTOC.

In order to study the relation between
infinite-temperature OTOC and the QPT in Rabi model, we numerically calculate the
time evolution of infinite-temperature OTOC $\mathcal{F}_{\beta=0}(t)$ with the frequency ratio $\eta = 2^{20}$,
which are depicted in Fig.~\ref{inf_OTOC}(a). As shown in Fig.~\ref{inf_OTOC}(b) and (c), in the beginning of the time evolution of the OTOC $\mathcal{F}_{\beta=0}(t)$, it decays exponentially, i.e., $\mathcal{F}_{\beta=0}(t) \sim \exp(-\lambda_{L}t)$. Consequently, $\log(\mathcal{F}_{\beta=0}(t)) \propto -\lambda_{L} t$ and the slope is Lyapunov-like exponents $\lambda_{L}$. We chose the appropriate time interval to ensure that $\log(\mathcal{F}_{\beta=0}(t)) \propto -\lambda_{L} t$ is satisfied. For instance, we fit the data of $\log(\mathcal{F}_{\beta=0}(t))$ with $t\in[0.3\omega_{0},0.6\omega_{0}]$ for $0.5\leq g/g_{c}\leq1$ and with $t\in[0.1\omega_{0},0.4\omega_{0}]$ for $g/g_{c}\geq 1.1$. The relation between $\lambda_{L}$
and $g/g_{c}$ is shown in the inset of Fig.~\ref{inf_OTOC_3}, which indicates that the $\lambda_{L}$ with $\beta=0$ can not reveal the information of
QPT.

Next, the time average of infinite-temperature OTOC $\tilde{\mathcal{F}}_{\beta=0}=\frac{1}{t_{f}}\int_{0}^{t_{f}}dt \mathcal{F}_{\beta=0}(t)$
is discussed. In Fig.~\ref{inf_OTOC}(a), because of the integrability of Rabi model~\cite{Rabi_integral},
drastic oscillation can be observed in the dynamics of $\mathcal{F}_{\beta=0}(t)$. Therefore, the long-time average is required for
extracting the signature of QPT. We present the results of $\tilde{\mathcal{F}}_{\beta=0}$
with the time window $t_{f}\sim 500 \omega_{0}$ as a function of $g/g_{c}$ in Fig. \ref{inf_OTOC_3}. Here,
the value of $\tilde{\mathcal{F}}_{\beta=0}$ is normalized by $\tilde{\mathcal{F}}_{\beta=0}/\mathcal{F}_{\beta=0}(t=0) = \tilde{\mathcal{F}}_{\beta=0}/\text{Tr}[ W^{\dagger}(0)V^{\dagger}(0)W(0)V(0)]$. Remarkably,
there is a local extreme point of $\tilde{\mathcal{F}}_{\beta=0}$ close to the critical point $g/g_{c}=1$, which
might be regarded as a signature of QPT and should be further ensured by studying the scaling behaviors of
the $\tilde{\mathcal{F}}_{\beta=0}$. The comparison between $\lambda_{L}$ extracted from the short-time behaviors of $\mathcal{F}_{\beta=0}(t)$ and $\tilde{\mathcal{F}}_{\beta=0}$ indicates the necessity of longer-time average of OTOC for the detection of QPT.

As a side remark, a recent work, where the two-point correlators (TPCs) in the Sachdev-Ye-Kitaev model and disordered XXZ model are studied, has revealed that the TPCs can characterize quantum chaos and be regarded as an alternative of OTOCs~\cite{two_corr}. In Fig. \ref{inf_OTOC_3}, we show that the infinite-temperature TPCs defined as $f_{\beta=0} = \text{Tr}[W(t)V(0)]$ with $\hat{W}=\hat{V}=a^{\dagger}a$ can also detect the QPT in the Rabi model via its time average $\tilde{f}_{\beta=0}(t)=\frac{1}{t_{f}}\int_{0}^{t_{f}}dt f_{\beta=0}(t)$ with $t_{f}\sim 500\omega_{0}$. However, the infinite-temperature OTOC is more sensitive to $g/g_{c}$ and the critical behavior is more distinct because the value of $\tilde{\mathcal{F}}_{\beta=0}$ at the critical point $g/g_{c}=1$ is smaller than that of $\tilde{f}_{\beta=0}(t)$.

In order to intuitively understand the normalization of $\tilde{\mathcal{F}}_{\beta=0}$, here we present how the value of infinite-temperature OTOC without normalization scales with the cavity photon number $n$. As shown in Fig. \ref{photon}(a), with the increase of $n$, both the $\mathcal{F}_{\beta=0}(t)$ at $t=0$ and $t=50$ increase with the form of $\mathcal{F}_{\beta=0}(t) \sim n^{4}$. In the limit of $n\rightarrow\infty$, the values of $\mathcal{F}_{\beta=0}(t)$ tend to infinite. Nevertheless, the ratio $\mathcal{F}_{\beta=0}(t=50)/\mathcal{F}_{\beta=0}(t=0)$ remains a finite value and smaller than 1 because $\mathcal{F}_{\beta=0}(t) < \mathcal{F}_{\beta=0}(t=0)$ for any $t>0$. Furthermore, we need to increase $n$ to suppress the influence of the $n$ effect. As depicted in Fig. \ref{photon}(b), the $\tilde{\mathcal{F}}_{\beta=0}$ is trivial for $n=20$. With the increase of $n$, the location of $\tilde{\mathcal{F}}_{\beta=0}$ minimum point tends to $g/g_{c}=1$. The locations of $\tilde{\mathcal{F}}_{\beta=0}$ minimum point for $n=60,80$ and $100$ are basically indistinguishable.
Therefore, the finite-$n$ effect on the behaviors of $\tilde{\mathcal{F}}_{\beta=0}$ can be neglected with $n\sim 80$.

\begin{figure}
	\centering
	\includegraphics[width=1\linewidth]{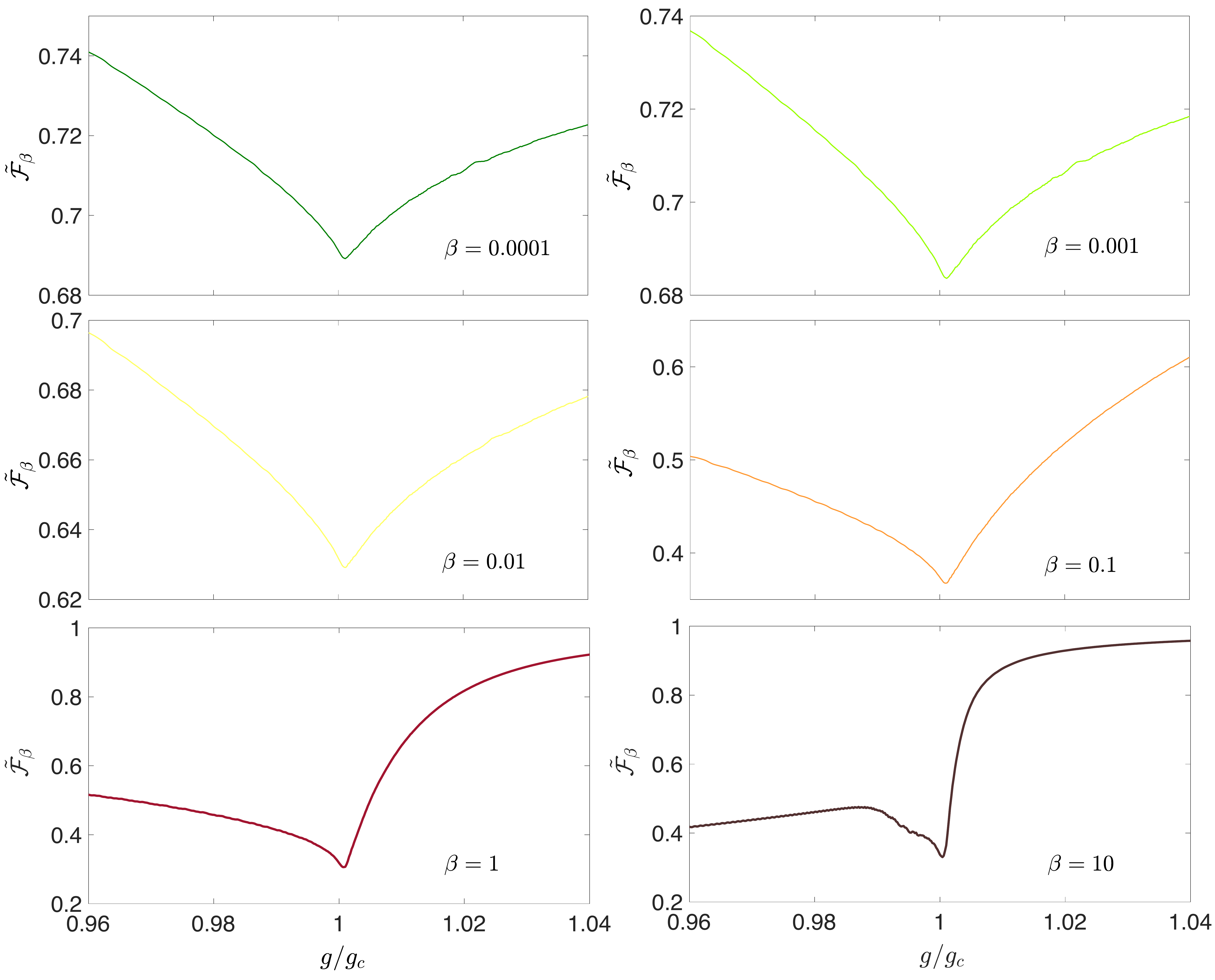}\\
	\caption{The time-averaged thermal OTOC $\tilde{\mathcal F}_{\beta}$ (rescaled by $\tilde{\mathcal F}_{\beta}/\tilde{\mathcal F}_{\beta}(0)$) in the Rabi model as a function of $g/g_{c}$ with $\eta=2^{20}$ and different finite values of $\beta$.}\label{finite_T}
\end{figure}

\begin{figure}
	\centering
	\includegraphics[width=1\linewidth]{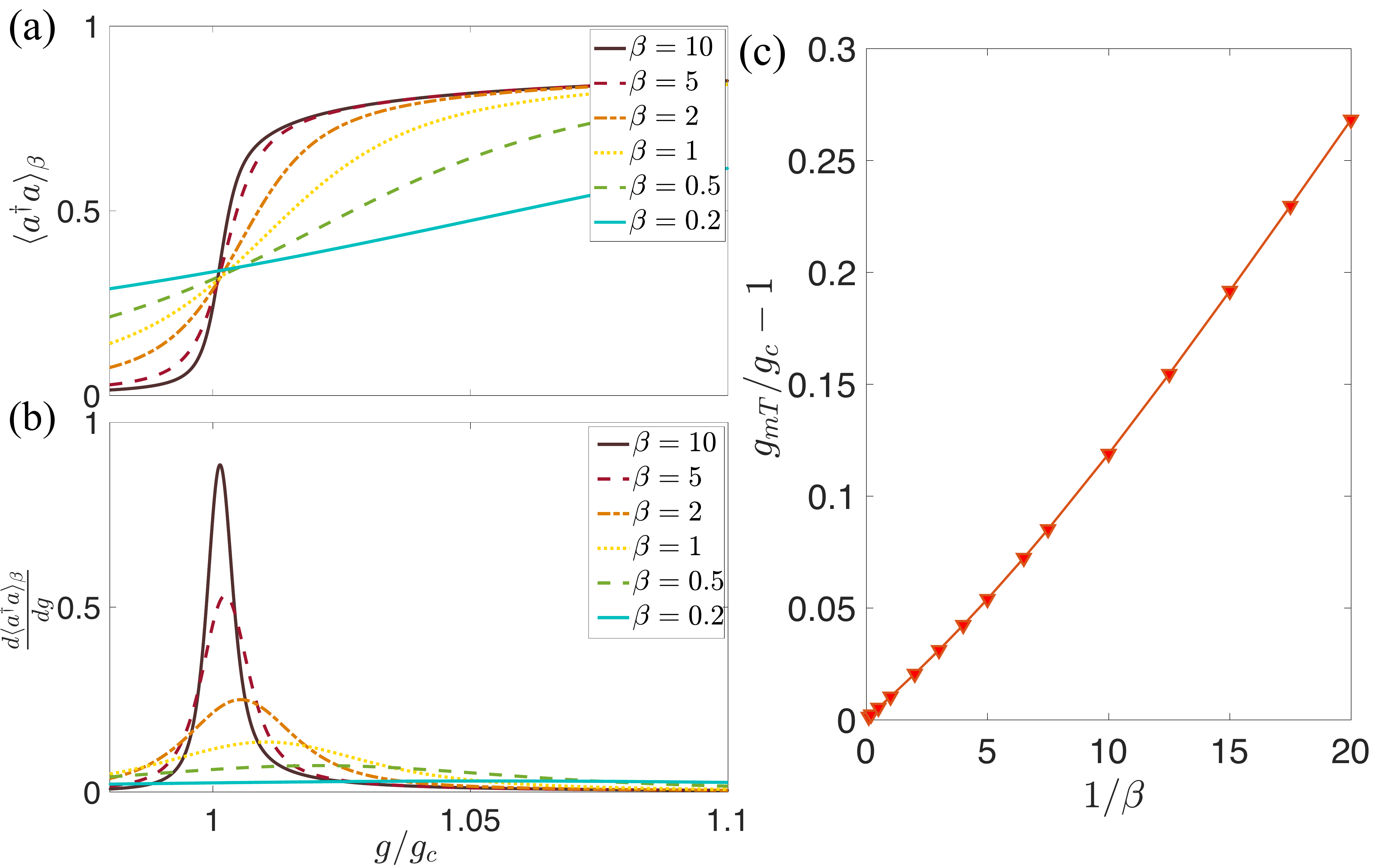}\\
	\caption{(a) The thermal average of order parameter $\langle a^{\dagger}a\rangle_{\beta}$ with different inverse-temperature $\beta=1/T$ in the Rabi model as a function of $g/g_{c}$ with $\eta=2^{20}$. (b) is Similar to (a) but for the susceptibility of $\langle a^{\dagger}a\rangle_{\beta}$ with respect to $g$, i.e., $\frac{d \langle a^{\dagger}a\rangle_{\beta}}{dg}$. (c) The location of maximum point of $\langle a^{\dagger}a\rangle_{\beta}$ as a function of temperature $T$.}\label{finite_T_add}
\end{figure}

It is well-known that the thermal fluctuations can influence the QPT that occurs at zero temperature in principle~\cite{book_QPT1}. For instance, the signatures of critical points dramatically change with the variation of $\beta$ in several systems such as the Dicke model or Ising model~\cite{Dicke_add2,TPT_add,TPT_add1,TPT_add2,TPT_add3}. However, as shown in Fig. \ref{finite_T}, it is remarkable that the local minimum points around $g/g_{c}=1$ of the time-averaged thermal OTOC $\tilde{\mathcal F}_{\beta}$ remain the same, and the capability of $\tilde{\mathcal F}_{\beta}$ in detecting the QPT is independent of $\beta$. Meanwhile, the shape of the $\tilde{\mathcal F}_{\beta}$ as a function of $g/g_{c}$ tends to the equilibrium OTOC depicted in Fig. \ref{eq_OTOC} with the increase of $\beta$. To compare the results of $\tilde{\mathcal F}_{\beta}$, we also calculate the thermal-averaged order parameter $\langle a^{\dagger}a\rangle_{\beta}$ and the results are shown in Fig. \ref{finite_T_add}. It can be obviously seen that the locations of $d \langle a^{\dagger}a\rangle_{\beta}/dg$ maximum point departure from the critical point $g_{c}$ at $T=0$. Thus, the signature of the QPT highlighted by the OTOC $\tilde{\mathcal F}_{\beta}$ is more robust against $\beta$ than the order paramater. Actually, besides the thermal average in the definition of the OTOCs, more importantly, $\mathcal{F}_{\beta}(t)$ probes the operator spread $\hat{W}(t)$ overlapping with the other operator $\hat{V}$, as a non-equilibrium property of the system, and could encode more information than the time-independent quantity $\langle a^{\dagger}a\rangle_{\beta}$.

\begin{figure}
  \centering
  \includegraphics[width=0.9\linewidth]{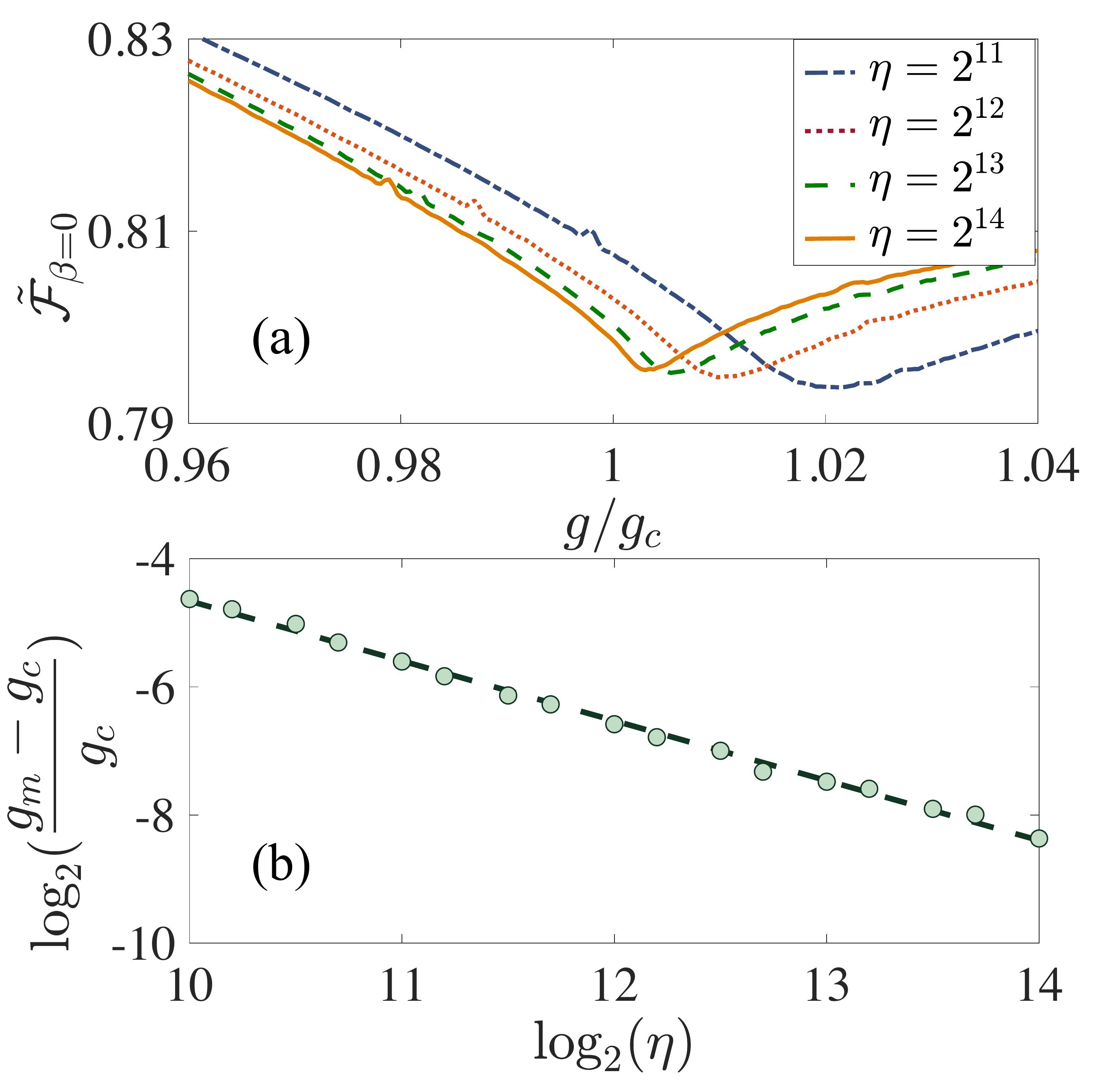}\\
  \caption{The time-averaged infinite-temperature OTOC $\tilde{\mathcal{F}}_{\beta=0}$ (rescaled by $\tilde{\mathcal{F}}_{\beta=0}/\mathcal{F}_{\beta=0}(0)$) as a function of $g/g_{c}$ with several values of $\eta$ in the Rabi model. (b) The dependence of $\log_{2}(\frac{g_{m}-g_{c}}{g_{c}})$ and $\log_{2}(\eta)$ where $g_{m}/g_{c}$ denotes the location of $\tilde{\mathcal{F}}_{\beta=0}$ minimum point. The green dashed line is the linear fitting $\log_{2}(\frac{g_{m}-g_{c}}{g_{c}})\sim -0.952 \log_{2}(\eta)$.}\label{scaling_rabi}
\end{figure}

Although the Rabi model can be realized in many experimental platforms~\cite{Rabi_exp1,Rabi_exp2,Rabi_exp3,Rabi_exp4,Rabi_exp5},
the limit $\eta\rightarrow \infty$ is unattainable in quantum simulation experiments.
This motivates us to study the finite-$\eta$ behaviors of $\tilde{\mathcal{F}}_{\beta=0}$ near the critical points.
In Fig. \ref{scaling_rabi}(a), we observe that the location of $\tilde{\mathcal{F}}_{\beta=0}$ minimum point $g_{m}/g_{c}$ is dependent on $\eta$ and
tends to $g_{m}/g_{c}=1$ with the increase of $\eta$. As shown in Fig. \ref{scaling_rabi}(b),
we obtain the scaling law
\begin{eqnarray}
\log(\frac{g_{m}-g_{c}}{g_{c}}) \sim -k\log(\eta),
\label{scaling_law}
\end{eqnarray}
where the positive number $k$ is the critical exponent of $\tilde{\mathcal{F}}_{\beta=0}$.
The scaling law is of great importance. Firstly, we can obtain the location of the critical
point in the limit $\eta\rightarrow\infty$
from the data of $\tilde{\mathcal{F}}_{\beta=0}$ with finite $\eta$. Secondly, we can decode the information of
universality class from the value of $k$. Because of the scaling law Eq. (\ref{scaling_law}), the emergence of a dip in $\tilde{\mathcal{F}}_{\beta=0}$ shown in Fig. \ref{inf_OTOC}(b) is a
signature of QPT.

\subsection{\label{sec:level2}OTOCs in the Dicke model}

\begin{figure}
  \centering
  \includegraphics[width=1\linewidth]{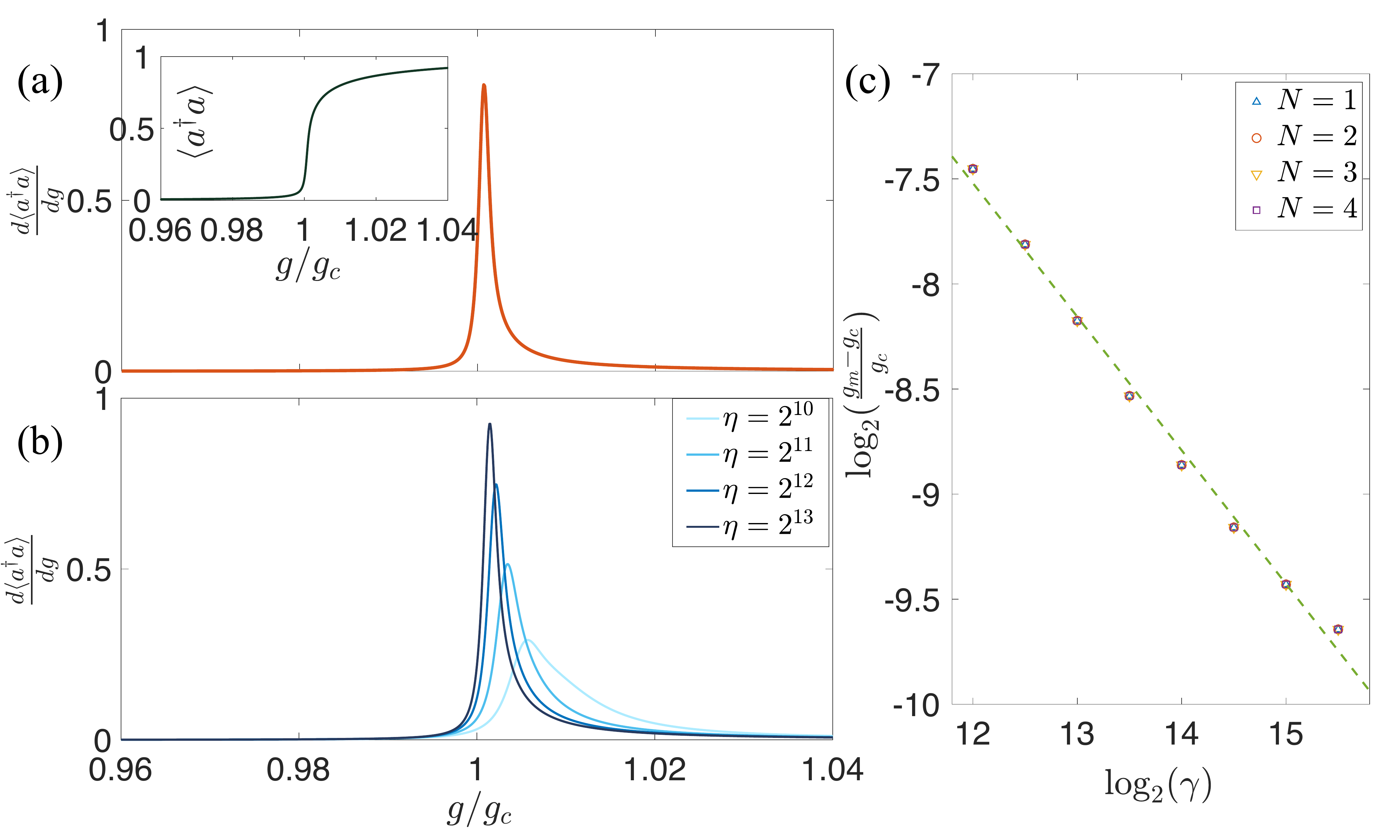}\\
  \caption{ (a) The value of $\frac{d\langle a^{\dagger}a\rangle}{dg}$ as a function of $g/g_{c}$. The inset shows the value of $\langle a^{\dagger}a\rangle$ (rescaled by the dimension of operator $a^{\dagger}a$ $n$) as a function of $g/g_{c}$. (b) The value of $\frac{d\langle a^{\dagger}a\rangle}{dg}$ as a function of $g/g_{c}$ for the Dicke model with $N=4$ and $\Omega/\omega_{0}=2^{10},2^{11},2^{12},2^{13}$. (c) The dependence of $\frac{g_{m}-g_{c}}{g_{c}}$, where $g_{m}$ is the location of maximum point of $\frac{d\langle a^{\dagger}a\rangle}{dg}$, and $\gamma=\Omega N/\omega_{0}$ with $N=1,2,4,8$. The dashed line is linear fitting $\log_{2}[(g_{m}-g_{c})/g_{c}]\sim -0.649\log_{2}(\gamma)$.}\label{eq_Dicke}
\end{figure}

The Rabi model can be generalized to the Dicke model by replacing single two-level atom with many two-level atoms.
The Hamiltonian of the Dicke model is
\begin{eqnarray}
\mathcal{H}^{N}_{\text{Dicke}} = \omega_{0}a^{\dagger}a+\Omega J_{z}+\frac{g}{\sqrt{2j}}(a^{\dagger}+a)(J_{+}+J_{-}),
\label{hamiltonian2}
\end{eqnarray}
where $N=2j$ is the number of atoms. Due to the permutation symmetry, $J_{z,\pm}=\sum_{i=1}^{2j}\frac{1}{2}\sigma^{i}_{z,\pm}$ can be defined as the collective spin operators with length $j$~\cite{Permutation1} .

State-of-art technology allows us to manipulate and obtain the dynamical result from the Dicke model with relatively small $N$~\cite{add_SR}.
Thus, it is worthy of studying further if there exists the second-order QPT in the Dicke model with extremely large $\eta=\Omega/\omega_{0}$ in the case of a small number of atoms. In the inset of Fig. \ref{eq_Dicke} (a), we show that $\langle a^{\dagger}a\rangle = \langle \psi | a^{\dagger}a|\psi\rangle$ with the ground state $|\psi\rangle$ is zero in the normal phase and becomes finite value in the superradiant phase, which is similar to $\langle a^{\dagger}a\rangle$ in the Rabi model~\cite{Rabi1}. Hence, a second-order QPT still exists in the few-body Dicke model in the limit $\eta\rightarrow \infty$ and $a^{\dagger} a$ is also a bona fide order parameter. As depicted in Fig. \ref{eq_Dicke} (a), there is a nonanalytical behavior for $d\langle a^{\dagger}a\rangle/dg$ diverges at the critical point $g/g_{c}=1$. In Ref.~\cite{Dicke2}, it has been revealed that the behaviors of order parameter susceptibility for the QPT of the Dicke model tend to divergence with the increase of $\eta$ or $N$, which is consistent with the results presented in Fig. \ref{eq_Dicke}(b). According to the dependence of $g_{m}$, as the location of the maximum point of $d\langle a^{\dagger}a\rangle/ d g$, and $\eta$, we can obtain a quantitative relation $\log(\frac{g_{m}-g_{c}}{g_{c}})\propto \log(\eta)$ whose formulation is the same as Eq. (\ref{scaling_law}). Furthermore, we demonstrate the relation for different $N$. As shown in Fig. \ref{eq_Dicke}(c), with the fixed $\gamma=\eta N=\Omega N/\omega_{0}$, $g_{c}$ is independent of $N$, indicating that $\gamma$ plays a more fundamental role in the studied QPTs and the generalized relation is $\log(\frac{g_{m}-g_{c}}{g_{c}})\propto \log(\gamma)$. In fact, similar to the results shown in Fig. \ref{eq_Dicke}(c), the scaling functions studied in Ref.~\cite{Rabi2} can also be generalized to $N>1$.

It is noted that various quantum optics experiments pay attention to the measurement of both adiabatic~\cite{Dicke_exp1,Dicke_exp_add1,Rabi_exp2} and quench dynamics~\cite{sc_add_exp} of order parameter, trying to observe similar signatures shown in the inset of Fig. \ref{eq_Dicke}(a). However, for the purpose of experimentally study the scaling law Eq. (\ref{scaling_law}), the cusp-like character of $\tilde{\mathcal{F}}_{\beta}$ in Fig. \ref{scaling_rabi}(a) may be more convenient than the order parameter $\langle a^{\dagger}a\rangle$ because $\tilde{\mathcal{F}}_{\beta}$ can directly pinpoint the critical point of the QPT, while we should make a derivative of $\langle a^{\dagger}a\rangle$, as a requirement for higher accuracy of experimental data, to obtain the similar cusp-like character, see Fig. \ref{scaling_rabi}(b).

%In addition, we realize that besides the quantum critical phenomenon, the chaotic-integrable transition also exists in the Dicke model~\cite{chaos1,chaos_Dicke_add1,chaos_Dicke_add2}.
%The chaos can be identified by the $\mathcal{F}_{\beta=0}(t)$ via monotonous decay with the increase of time~\cite{mbl4}. However, as shown in Fig. \ref{dynamic_Dicke}, the obvious fluctuation in the time evolution of $\mathcal{F}_{\beta=0}(t)$ for the Dicke model with $N=4$ and $\eta=2^{20}$ near the critical point can be observed, which indicates that the system is nonergodic in this condition.
%Therefore, the signature of QPT revealed via $\mathcal{F}_{\beta=0}$
%is not influenced by the chaotic-integrable transition.
%It has been shown that there is a transition from nonergodicity to ergodicity
%with the increase of coupling strength $g$ in the Dicke model with $\eta=1$~\cite{chaos1}.
%Whereas, the nonergodicity in the few-body Dicke model with large $\eta$ remains an intriguing open question.

%
\begin{figure}
  \centering
  \includegraphics[width=0.9\linewidth]{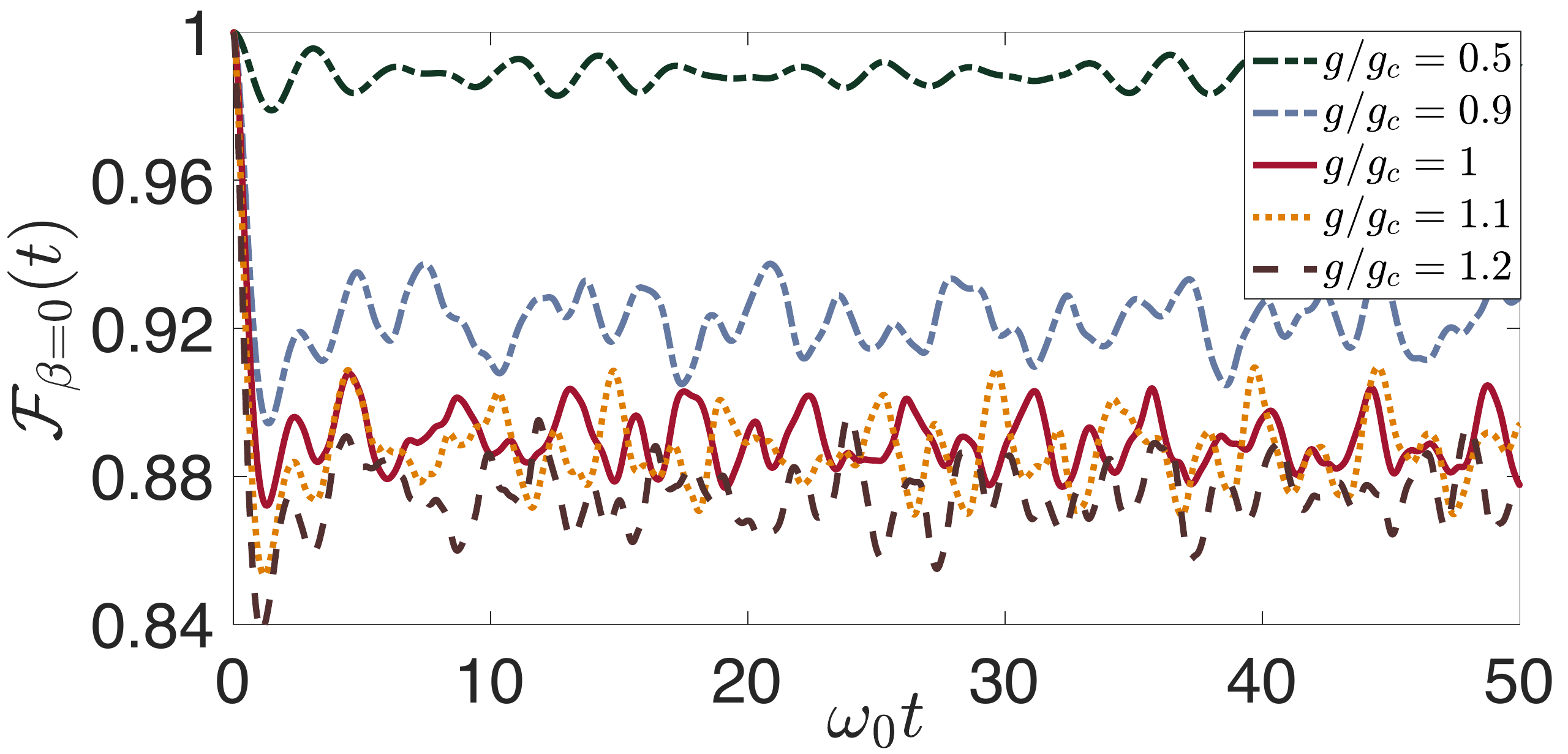}\\
  \caption{Time evolution for the infinite-temperature OTOC $\mathcal{F}_{\beta=0}(t)$ in the Dicke model with $\eta=2^{20}$, $N=4$, and $g/g_{c}=0.5,0.9,1,1.1$ and $1.2$.}\label{dynamic_Dicke}
\end{figure}
\begin{figure}
  \centering
  \includegraphics[width=1\linewidth]{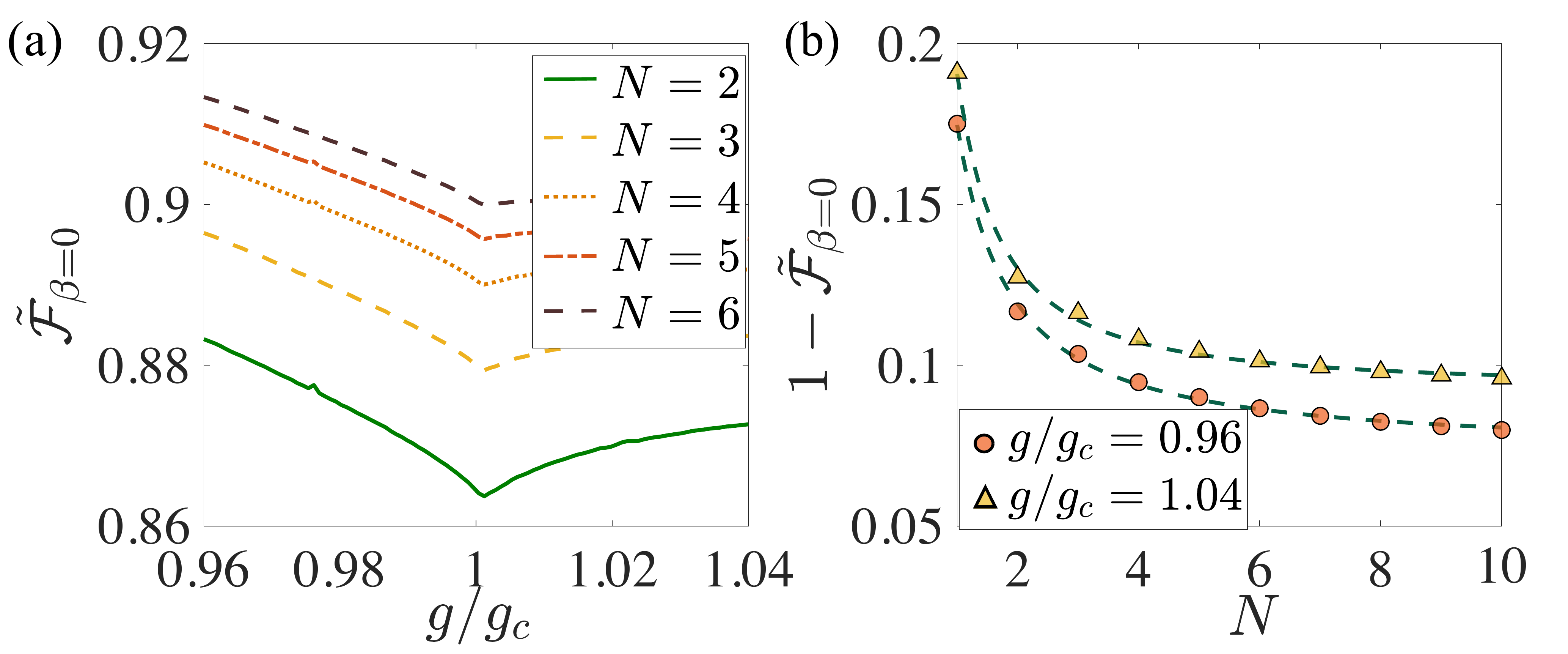}\\
  \caption{(a) $\tilde{\mathcal{F}}_{\beta=0}$ (rescaled by $\tilde{\mathcal{F}}_{\beta=0}/\mathcal{F}_{\beta=0}(0)$) as a function of $g/g_{c}$ in the Dicke model with different number of atoms $N$ and $\eta=2^{20}$. (b) The value of $1-\tilde{\mathcal{F}}_{\beta=0}$ as a function of $N$. The dashed lines are fitting curves with the formulation of Eq. (\ref{block_scaling}).}\label{otoc_functions_Dicke}
\end{figure}

We then explore the linkage between the QPTs in the Dicke model and the infinite-temperature OTOC, as well as the related scaling laws. After calculating the dynamics of infinite-temperature OTOC shown in Fig. \ref{dynamic_Dicke}, 
we can obtain the time-averaged infinite-temperature OTOC $\tilde{\mathcal{F}}_{\beta=0}$ as a function of $g/g_{c}$ in the Dicke model with different $N$ and $\eta=2^{20}$. The results are depicted in Fig. \ref{otoc_functions_Dicke}(a). Similar to the behaviors of $\tilde{\mathcal{F}}_{\beta=0}$ in the Rabi model, we can also observe the local minimum point of $\tilde{\mathcal{F}}_{\beta=0}$ near $g/g_{c}=1$ in the Dicke model. Consequently, the infinite-temperature OTOC is capable of identifying the critical point of QPT in the Dicke model with large $\eta$.

\begin{figure}
  \centering
  \includegraphics[width=1\linewidth]{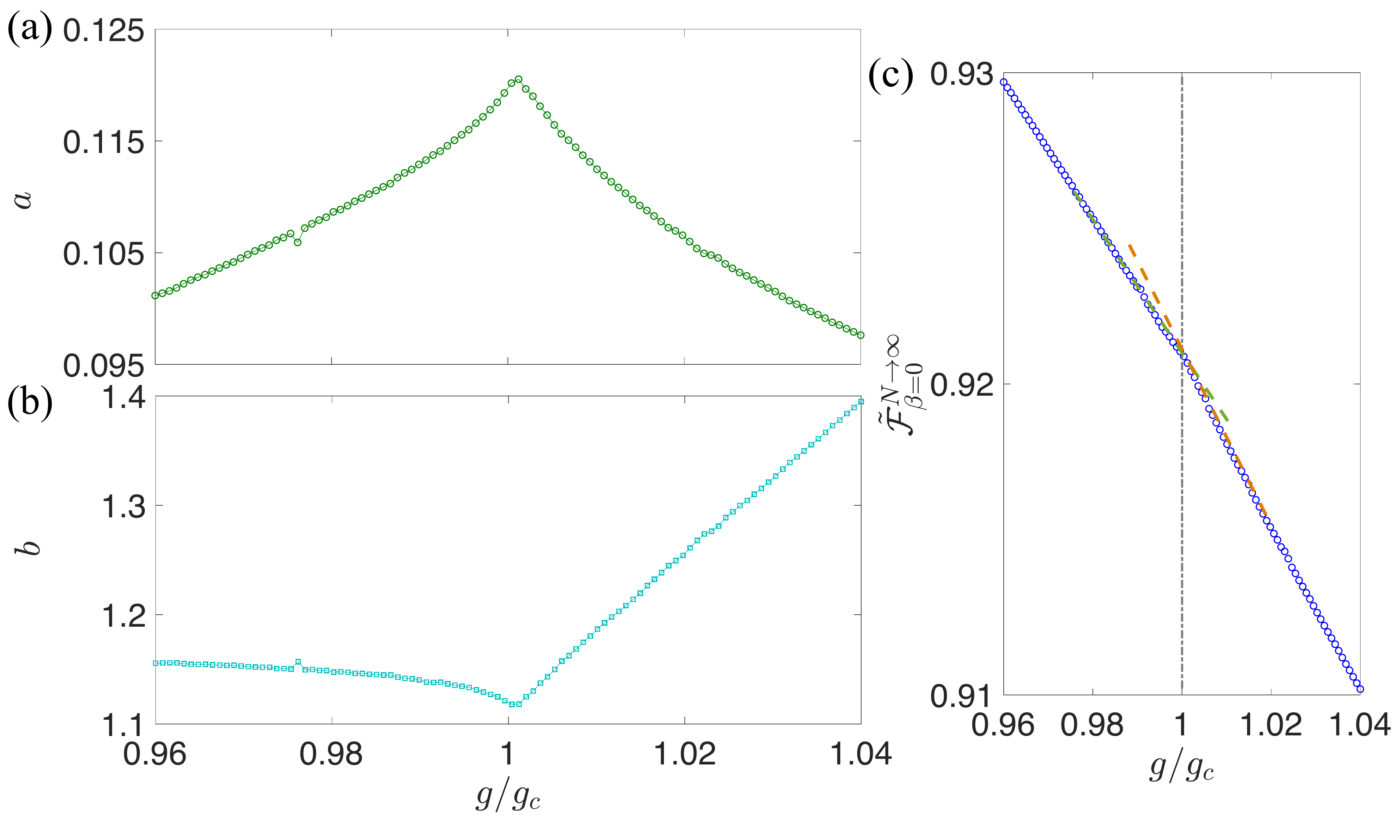}\\
  \caption{The coefficient (a) $a$, (b) $b$, and (c) $c$ in Eq. (\ref{block_scaling}) as a function of $g/g_{c}$.}\label{para}
\end{figure}

Moreover, the finite-$N$ effect on the value of $\tilde{\mathcal{F}}_{\beta=0}$ is also studied. As shown in Fig. \ref{otoc_functions_Dicke}(b), with the increase of $N$, the value of $1-\tilde{\mathcal{F}}_{\beta=0}$ at critical point decays as a power law,
\begin{eqnarray}
1-\tilde{\mathcal{F}}_{\beta=0} = aN^{-b} + c.
\label{block_scaling}
\end{eqnarray}
The coefficient $a$, $b$ as a function of $g/g_{c}$ is presented in Fig. \ref{para}(a) and (b) respectively, which indicates that the coefficients can pinpoint the critical point. Moreover, $\tilde{\mathcal{F}}^{N\rightarrow\infty}_{\beta=0} = 1-c$ can be regarded as the value of $\tilde{\mathcal{F}}_{\beta=0}$
in the thermodynamic limit $N\rightarrow\infty$. The result of the coefficient $c$ depicted in Fig. \ref{para}(c) suggests that $\tilde{\mathcal{F}}_{\beta=0}$ can still characterize QPT in this limit.

\begin{figure}
  \centering
  \includegraphics[width=1\linewidth]{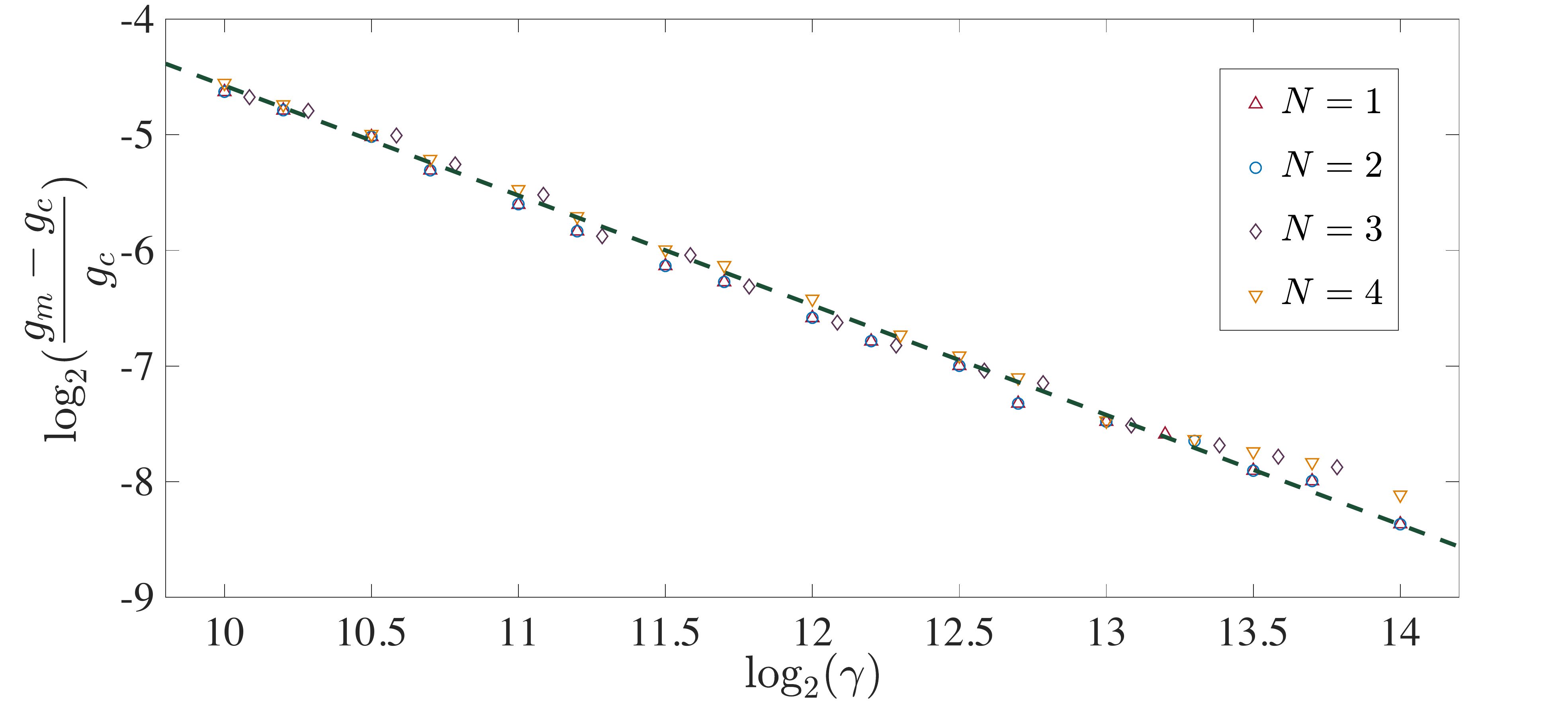}\\
  \caption{The dependence of $\log_{2}(\frac{g_{m}-g_{c}}{g_{c}})$ and $\log_{2}(\gamma)$ where $g_{m}/g_{c}$ denotes the location of $\tilde{\mathcal{F}}_{\beta=0}$ minimum point in the Dicke model with $N=1,2,3,4$. The dashed lines are linear fittings with the formulation $\log_{2}(\frac{g_{m}-g_{c}}{g_{c}}) \sim -0.949 \log_{2}(\gamma)$.}\label{Dicke_scaling}
\end{figure}

It is now recognized that in the Dicke model, the quantity $\gamma = \Omega N/\omega_{0}$ dominates the criticality~\cite{Rabi2}.
We proceed to explore the the finite-$\gamma$ scaling law of $\tilde{\mathcal{F}}_{\beta=0}$,
which can be regarded as the
universal properties of all the finite-$N$ Dicke model.
As shown in Fig. \ref{Dicke_scaling}, the formulation of scaling law is
\begin{eqnarray}
\log(\frac{g_{m}-g_{c}}{g_{c}}) \sim -\kappa\log(\gamma),
\label{scaling_law_size}
\end{eqnarray}
where $g_{m}/g_{c}$ denotes the location of $\tilde{\mathcal{F}}_{\beta=0}$ minimum point.
The scaling exponents $\kappa$ and $k$ play crucial roles in the universality of QPTs.
Comparing the results in Fig. \ref{scaling_rabi} (b) and Fig. \ref{Dicke_scaling}, it is revealed that $\kappa \simeq k$ and
the QPTs in the Rabi and Dicke model belong to the same universality class. Consequently, the scaling laws of $\tilde{\mathcal{F}}_{\beta=0}$
are not only helpful for estimating the location of critical point with finite-$\eta$ and finite-$N$,
but also determining the universal class of QPTs. We also note that
the formulation of scaling law Eq. (\ref{scaling_law_size}) is consistent with the results of
the ground state energy in few-body Dicke model with finite-$\eta$ shown in Fig. \ref{eq_Dicke} (c).

\section{Conclusion and outlook}

We have presented numerical evidence
showing that the time-averaged infinite-temperature OTOC $\tilde{\mathcal{F}}_{\beta=0}$ displays a minimum around the critical points
in both the Rabi and Dicke model.
In this way, the infinite-temperature OTOC dynamically diagnoses the QPT between normal phase and superradiant phase.
In finite frequency system, the QPT does not happen at the exact critical point calculated from the infinite frequency system, which will give a finite-frequency scaling law.
The scaling laws of $\tilde{\mathcal{F}}_{\beta=0}$ are identified to confirm the occurrence of QPTs.
In the Rabi model, we show that the $\tilde{\mathcal{F}}_{\beta=0}$ present
finite-frequency scaling behaviors close to the critical point. In the Dicke model, we also demonstrate
the scaling behaviors of $\tilde{\mathcal{F}}_{\beta=0}$
with finite values of $\gamma = \Omega N/\omega_{0}$.
The scaling laws are of great significance for phase transition.
Based on the scaling laws,
we characterize the QPTs in the limit $\gamma \rightarrow \infty$ from the data of OTOC with finite values of $\gamma$.
Moreover, the universality class of QPTs has been identified by the scaling exponent obtained from the scaling laws.
We find that the QPTs of the Dicke models with different number of atom(s) $N$ belong to the same universality class, i.e.,
the $N$-body Dicke models ($N=1$ for the Rabi model), share a common scaling law.

This work may shed light on the characterization of QPT
via the dynamics of R\'{e}nyi entropy (RE) based on the OTOC-RE theorem~\cite{mbl5}.
Actually, the quenched dynamics of R\'{e}nyi entropy has been widely studied in many-body systems theoretically~\cite{renyi1} and experimentally~\cite{renyi2,renyi3}. Nevertheless, the quenched dynamics of R\'{e}nyi entropy
in cavity-atom interaction systems and its linkage between
the QPTs and chaos in the systems are worthwhile to study further.
On the other hand, this work could also enlighten the investigations of thermal averaged OTOCs in condensed-matter systems~\cite{OTOC_XXZ,OTOC_topo,OTOC_qpt_add} and other quantum optics systems such as driven Tavis-Cummings model~\cite{sc_add_exp}.
With the rapid development of quantum simulation~\cite{Dicke_exp1,Dicke_exp2,Dicke_exp3,Rabi_exp1,Rabi_exp2,Rabi_exp3,Rabi_exp4,Rabi_exp5,decoherence1,decoherence2,decoherence3} and general measurement protocol of OTOCs~\cite{random_measure},
we believe that our results may provide an experimentally feasible approach to detect equilibrium quantum critical points
and find the universal properties of few-body systems in quantum region.

\section*{Acknowledgments}

J.-Q Cai thanks Ming Gong for
helpful discussions. We are grateful to the numerical
packages of the Python, Numpy, Scipy and QuTiP~\cite{qutip}
open source projects. Y. H. was partially supported by National Natural Science Foundation of China (Grant No. 11774114). H. F. was partially supported by Ministry of Technology of China (Grants No. 2016YFA0302104 and No. 2016YFA0300600), National Natural Science Foundation of China (Grant No. 11774406), and Chinese Academy of Science (Grant No. XDPB0803).

\section*{Conflict of Interest}

The authors declare no conflict of interest.

\section*{Keywords}

out-of-time-order correlators, quantum phase transtions, Rabi model, Dicke model 

%%%%%%%%%%%%%%%%%%%%%%%%%%%%%%%%%%%

%%%%%%%%%%%%%%%%%%%%%%%%%%%%%%%

\end{document}